\documentclass[aps,prd,twocolumn,superscriptaddress,preprintnumbers,floatfix,nofootinbib,notitlepage,showkeys,showpacs]{revtex4-1}

\usepackage[utf8]{inputenc}

\usepackage{graphicx}
\usepackage{hyperref}
\usepackage{latexsym}
\usepackage{amsmath}
\usepackage{amssymb}
\usepackage{bbm}
\usepackage{ulem}
\usepackage{pdfsync}
\usepackage{epsfig}
\usepackage{epstopdf}
\usepackage{subfigure}
\usepackage{xcolor}
\usepackage{comment}
\usepackage{slashed}
\usepackage{upgreek}

\long\def\exclude#1{}

\newcommand{\beq}{\begin{equation}}
\newcommand{\eeq}{\end{equation}}

\usepackage{scalerel,tikz}
\usetikzlibrary{svg.path}
\definecolor{orcidlogocol}{HTML}{A6CE39}
\tikzset{orcidlogo/.pic={
 \fill[orcidlogocol] svg{M256,128c0,70.7-57.3,128-128,128C57.3,256,0,198.7,0,128C0,57.3,57.3,0,128,0C198.7,0,256,57.3,256,128z};
 \fill[white] svg{M86.3,186.2H70.9V79.1h15.4v48.4V186.2z}
 svg{M108.9,79.1h41.6c39.6,0,57,28.3,57,53.6c0,27.5-21.5,53.6-56.8,53.6h-41.8V79.1z M124.3,172.4h24.5c34.9,0,42.9-26.5,42.9-39.7c0-21.5-13.7-39.7-43.7-39.7h-23.7V172.4z}
 svg{M88.7,56.8c0,5.5-4.5,10.1-10.1,10.1c-5.6,0-10.1-4.6-10.1-10.1c0-5.6,4.5-10.1,10.1-10.1C84.2,46.7,88.7,51.3,88.7,56.8z};
}}
\newcommand\orcidicon[1]{\href{https://orcid.org/#1}{\mbox{\scalerel*{
\begin{tikzpicture}[yscale=-1,transform shape]
\pic{orcidlogo};
\end{tikzpicture}
}{|}}}}

\allowdisplaybreaks

\bibpunct{[}{]}{,}{n}{}{,}

\newcommand{\n}{\hat{\pmb{n}}}
\newcommand{\kv}{\pmb{k}}
\newcommand{\xv}{\pmb{x}}

\begin{document}

\title{Removal of interloper contamination to line-intensity maps using correlations with ancillary tracers of the large-scale structure}
\author{Jos\'e Luis Bernal~\orcidicon{0000-0002-0961-4653}}
\affiliation{Instituto de Física de Cantabria (IFCA), CSIC-Univ. de Cantabria, Avda. de los Castros s/n, E-39005 Santander, Spain}
\author{Ant\'on Baleato Lizancos~\orcidicon{0000-0002-0232-6480}}
\affiliation{Berkeley Center for Cosmological Physics, UC Berkeley, CA 94720, USA}
\affiliation{Department of Physics, University of California, Berkeley, CA 94720, USA}
\affiliation{Lawrence Berkeley National Laboratory, One Cyclotron Road, Berkeley, CA 94720, USA}


\begin{abstract}
Line-intensity mapping (LIM) offers an approach to obtain three-dimensional maps of the large-scale structure by collecting the aggregate emission from all emitters along the line of sight. The procedure hinges on reconstructing the radial positions of sources by relating the observed frequency to the rest-frame frequency of a target emission line. However, this step is hindered by `interloper-line' emission from different cosmological volumes that redshifts into the same observed frequency. In this work, we propose a model-independent technique to remove the contamination of line interlopers using their statistical correlation with external tracers of the large-scale structure, and identify the weights that minimize the variance of the cleaned field. 
Furthermore, we derive expressions for the resulting power spectra after applying our cleaning procedure, and validate them against simulations. We find that the cleaning performance improves as the correlation between the line interlopers and the external tracer increases, resulting in a gain in the signal-to-noise ratio of up to a factor 6 (2) for the auto- (cross-)power spectrum in idealized scenarios. This approach has the advantage of being model-independent, and is highly complementary to other techniques, as it removes interloper large-scale clustering modes instead of individually masking the brightest sources of contamination.

\end{abstract}

\maketitle

\section{Introduction}
\label{sec:intro}
Line-intensity mapping (LIM) is an observational technique that employs flux integrated over cosmological scales to obtain three-dimensional maps of the Universe, recovering the line-of-sight information by targeting bright spectral lines with known rest-frame frequency~\cite{Liu:2019awk, Bernal:2022jap}. This approach grants access to the aggregate emission of otherwise undetectable faint sources and offers sensitivity to the astrophysical processes triggering the line emission. Therefore, LIM provides a powerful avenue to trace the large-scale structure of the Universe and galaxy formation and evolution at high redshifts, especially in regimes close to the high-noise or high-confusion limits~\cite{Cheng:2018hox, Schaan:2021hhy}. 

The LIM experimental landscape is growing rapidly, with an increasing number of facilities currently observing~\cite{vanHaarlem:2013dsa, Bandura:2014gwa, DeBoer:2016tnn, MeerKLASS:2017vgf, Cleary:2021dsp, CONCERTO:2020ahk, Gebhardt:2021vfo} and many others that will start in the forthcoming years~\cite{CCAT-Prime:2021lly, Sun:2020mco, Switzer:2021jeg, 2020arXiv200914340V, Dore:2014cca, Koopmans:2015sua, Newburgh:2016mwi}. These experiments are expected to confirm and increase the significance of preliminary detections (see e.g., Refs.~\cite{Chang:2007xk, Keating:2016pka, Yang:2019eoj, Keating:2020wlx, Wolz:2021ofa, Niemeyer:2022vrt, Cunnington:2022uzo, Niemeyer:2022arn, Paul:2023yrr}), scale up the size and sensitivity of the current pathfinder stage, and bridge the low-redshift and recombination epochs of the Universe probed by current experiments (see e.g., Refs.~\cite{MoradinezhadDizgah:2018lac, Bernal:2019gfq, Munoz:2019fkt, Sato-Polito:2020qpc, Sato-Polito:2020cil, MoradinezhadDizgah:2021upg}).

Nonetheless, some of the particularities that make LIM so appealing also feature among its main challenges. A critical one among these stems from the fact that, by using the integrated flux, LIM observations also collect contamination from any unwanted emission along the line of sight. This contamination includes Galactic foregrounds, continuum extragalactic emission, and line interlopers ---i.e., emission from lines other than the target that redshifts into the same observed frequency. How problematic each of these foregrounds is depends on the frequency of the observations and the spectral line targeted. For instance, there is no relevant emission line close in frequency to the 21 cm line, hence line interlopers do not affect 21-cm line-intensity mapping observations at all. On the contrary, the spectrum at higher frequencies presents several bright emission lines. Therefore, line-intensity mapping experiments targeting these lines must account also for line-interloper contamination.

The contribution of 
Galactic foregrounds and continuum extragalactic emission is expected to be smooth enough in frequency to be mitigated with component separation (see e.g., Ref.~\cite{Carucci_inprep}), though at the cost of losing the longest-wavelength modes along the line of sight~\cite{Switzer:2018tel, Cunnington:2023jpq}. On the other hand, removing line-interloper contamination ---noisier in frequency than continuum emission and more similar to the signal of interest--- is a greater challenge. 

Although the intensity fluctuations of the interloper lines are uncorrelated with those of the target line,\footnote{Most of the commonly-targeted spectral lines are chosen to be bright and to have no other lines close to them in frequency to avoid line confusion.} they nevertheless contribute to the total intensity measured and suffer from strong projection effects due to the confusion in redshift when projecting the position on the sky and observed frequency onto three-dimensional, Cartesian positions. Foreground line interlopers are typically brighter than the contribution of interest, even though the target spectral line may be intrinsically more intense, and can even dominate the measurements (see e.g, Ref.~\cite{Bethermin:2022lmd}). In fact, line interlopers may well be the most problematic source of contamination in LIM observations above radio and below optical frequencies. 

Line-interloper contamination can be avoided cross-correlating the line-intensity maps either with other spectral line at the same redshift or with other tracers of the large-scale structure (see e.g., Refs.~\cite{Lidz:2008ry, Silva:2014ira, COMAP:2018svn, Roy:2023pei}). The auto-power spectrum of the line of interest can be partially reconstructed from the cross-correlations of three tracers of the same density fluctuations at large scales~\cite{Beane:2018dzk}. However, if not mitigated, the contribution from line interlopers significantly increases the covariance of not only auto-correlations, but also cross-correlations. Although in this paper we will focus on line interlopers in the context of LIM, they can also be a challenge for certain galaxy surveys (see e.g., Refs.~\cite{Pullen:2015yba, GrasshornGebhardt:2018cho, Farrow:2021trw}), introducing catastrophic redshift errors and leading to impure galaxy catalogs. 

There are two main strategies to deal with interlopers: masking and modeling. Masking can be either blind or guided. Guided masking~\cite{2018ApJ...856..107S, 2021arXiv211105354S, VanCuyck:2023uli} uses external observations to locate galaxies that are expected to be bright in the interloper lines and masks the voxels\footnote{A voxel is a three-dimensional pixel ---a pixel within some frequency bin.} that they fall into. Therefore, this approach relies on some astrophysical model to guide the masking strategy, which may result in suboptimal cleaning. On the other hand, blind masking~\cite{2011JCAP...08..010V, Gong:2013xda, Breysse:2015baa} masks the brightest voxels (which are more prone to be dominated by foreground interloper emission) without requiring external observations, but incurs some (limited) information loss. Knowledge of the rest-frame frequency of the interlopers can also be used to separate interloper and target signals, either using spectral templates at the pixel level~\cite{Cheng:2020asz} or following machine learning approaches~\cite{2020MNRAS.496L..54M, Moriwaki:2020bpr, Moriwaki:2021yie}. Finally, the contribution from interlopers can be modeled and considered as part of the analysis (see e.g., Refs.~\cite{Gong:2013xda,Lidz:2016lub,Cheng:2016yvu, Cheng:2024nfy}), which allows for the search of exotic contributions~\cite{Bernal:2020lkd, Bernal:2021ylz}. Nonetheless, if the line interlopers are too bright they will hinder the study of the target line, despite efforts to model them.

In this work we propose an alternative approach that statistically minimizes the contamination from foreground line interlopers. Our approach is inspired by lensing-nulling techniques~\cite{huterer_nulling_2005}, especially those developed to undo the effects of lensing from the observed cosmic microwave background (CMB) anisotropies using external tracers~\cite{smith_delensing_2012, Sherwin:2015baa, baleato_lizancos_delensing_2022} or to minimize contributions from certain redshifts to the reconstructed CMB lensing convergence~\cite{McCarthy:2020dgq, Qu:2022xow, maniyar_new_2022, Lizancos:2023jpo}. Following a loose analogy, we will employ ancillary tracers of the large-scale structure ---typically fluctuations in the number density of galaxies measured by spectroscopic surveys--- that overlap in redshift with the foreground interlopers to remove contributions from the latter from LIM observations. Crucially, the ancillary tracers are first filtered in such a way that the variance of the \textit{cleaned} map is minimized.

We derive a prediction for the power spectrum of the cleaned map, and successfully validate it against log-normal simulations. This prediction can be calibrated directly from the data and is thus model independent. Since the performance of the cleaning is mostly driven by the extent of correlation between the interloper intensity fluctuations and the ancillary tracer we do the cleaning with, spectroscopic (or very-narrow-band photometric) galaxies are likely to be tracer best suited for this technique. 

A key benefit of our approach is that, by (partially) removing the exact realization of the interloper fluctuations present in the sky, the covariance of both LIM auto- and cross-spectra can be reduced. We provide estimates of the improvement factor in the signal-to-noise ratio on the detection of such summary statistics, which can be as high as a factor of 6 (2) for auto- (cross-) power spectra ---assuming Gaussian covariances. Given the strong dependence of the correlation coefficients ---and the severity of the projection effects--- on the specific redshifts and emission lines considered, we choose to remain as general as possible in this proof-of-concept study and defer in-detail investigation of specific examples to future work.

This paper is structured as follows. We review the theory of LIM signal, power spectrum and interloper contamination in Sec.~\ref{sec:lim}. The technique we propose to remove interloper contributions to the LIM power spectra is presented in Sec.~\ref{sec:nulling} and validated with log-normal simulations in Sec.~\ref{sec:validation}. The results and implications of the technique are discussed in Sec.~\ref{sec:discussion}. Finally, our conclusions are presented in Sec.~\ref{sec:conclusions}. Further details are relegated to the appendices, including a harmonic-space  version of our formalism (appendix~\ref{app:harmonic}) and an explanation of how to combine information from several tracers of the large-scale structure (appendix~\ref{app:multitracer}). 

\section{LIM: target signal and line interlopers}
\label{sec:lim}
Consider any line emission sourced from a position $\hat{\pmb{n}}$ on the sky at redshift $z$, with local specific luminosity density $\rho_L$ at rest-frame frequency $\nu$. Its specific intensity at $\nu_{\rm obs}=\nu/(1+z)$ is given by 
\begin{equation}   
\begin{split} 
    I_{\nu_{\rm obs}}(\hat{\pmb{n}}) = & \frac{c}{4\pi \nu_{\rm obs}(1+z) H(z)} \rho_L(\n\chi(z))  \\
    \equiv & X_{\rm LI}^{\nu_{\rm obs}}(z)\rho_L(\n\chi(z))\,,
\end{split}
\end{equation}
where $c$ is the speed of light, $H(z)$ is the Hubble parameter, $\chi$ is the comoving radial distance to redshift $z$, and we have defined $X_{\rm LI}^{\nu_{\rm obs}}$ to absorb all constants and quantities that depend only on redshift. Then, the total specific intensity observed at a given $\nu_{\rm obs}$ is the sum of all contributions that redshift into such frequency:
\begin{equation}
    I_{\nu_{\rm obs}}(\hat{\pmb{n}}) = \sum_i X_{\rm LI}^{\nu_{\rm obs}}(z_i)\rho_{L}^{(i)}(\n\chi(z_i))\,.
    \label{eq:Itot}
\end{equation}
In the equation above we only consider line emission (assuming that continuum sources have been already cleaned from observations), so that we can replace an integral over the line of sight with the sum of contributions from lines with rest-frame frequency $\nu^{(i)}=\nu_{\rm obs}(1+z^{(i)})$. Therefore, Eq.~\eqref{eq:Itot} accounts for the line of interest of the LIM survey as well as all the line interlopers that contaminate that signal. 

As is customary in LIM analyses, we will consider three-dimensional maps,\footnote{Note that there are also proposed methodologies employing angular statistics, see e.g., Ref.~\cite{Cheng:2024nfy}.} recovering line-of-sight information from the spectral resolution of the instrument assuming the rest-frame frequency $\nu_{\rm t}$ of the line of interest. This transformation means that two points that are separated by an angle $\vartheta$ on the sky and a redshift interval $\delta z$ (centered at $z$) are mapped onto separations
\begin{equation}
    r_\perp = D_{\rm M}(z)\vartheta\,, \qquad \text{and} \qquad r_\parallel = \frac{c\,\delta z}{H(z)}\,
    \label{eq:2d-to-3d}
\end{equation}
transverse to and along the line of sight, respectively, where $D_{\rm M}$ is the comoving angular diameter distance.\footnote{This transformation assumes the plane-parallel approximation, which introduces spurious artifacts in the computation of two-point statistics. We refer the interested reader to Ref.~\cite{Cunnington:2023aou} for a careful treatment in the context of LIM.} As mentioned above, the redshift $z_{\rm t}=\nu_{\rm t}/\nu_{\rm obs}-1$ is used to apply the transformation in Eq.~\eqref{eq:2d-to-3d}, so that contributions from line interlopers will be subject to projection effects, which introduce severe anisotropies in their fluctuations. For each interloper line, the \textit{measured} and \textit{true} components of the three-dimensional separations between two points are related by $r_j^{\rm true} = r_j^{\rm meas}q_j$, where the subscript $j$ stands for transverse to and parallel to the line of sight, and\footnote{These projection effects are qualitatively similar to the Alcock-Paczyński distortions~\cite{alcock_evolution_1979}. Since the latter ---sourced by differences between the assumed cosmology applied in Eq.~\eqref{eq:2d-to-3d} and the actual one--- are decoupled from the question we would like to investigate in this work, we neglect them and assume that our fiducial cosmology matches the true cosmology.} ~\cite{Lidz:2016lub}
\begin{equation}
    q_\perp^{(i)}=\frac{D_M(z_i)}{D_M(z_{\rm t})},\qquad \text{and} \qquad q_\parallel^{(i)} = \frac{(1+z_i)/H(z_i)}{(1+z_{\rm t})/H(z_{\rm t})}\,.
    \label{eq:qs_def}
\end{equation}

\begin{figure}[t]
\centering
\includegraphics[width=\columnwidth]{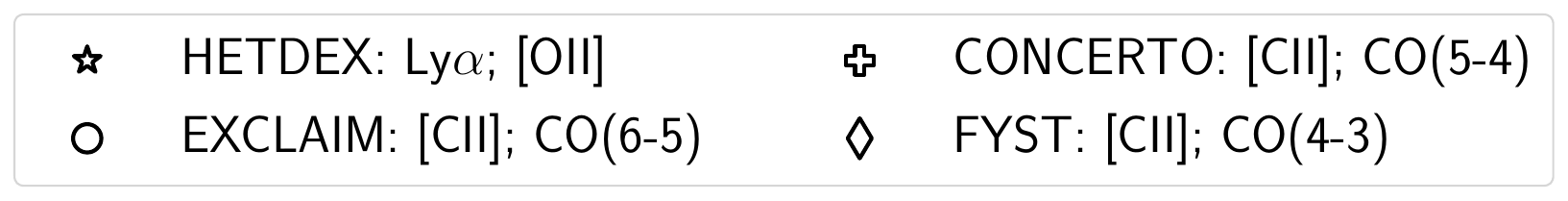}
\includegraphics[width=\columnwidth]{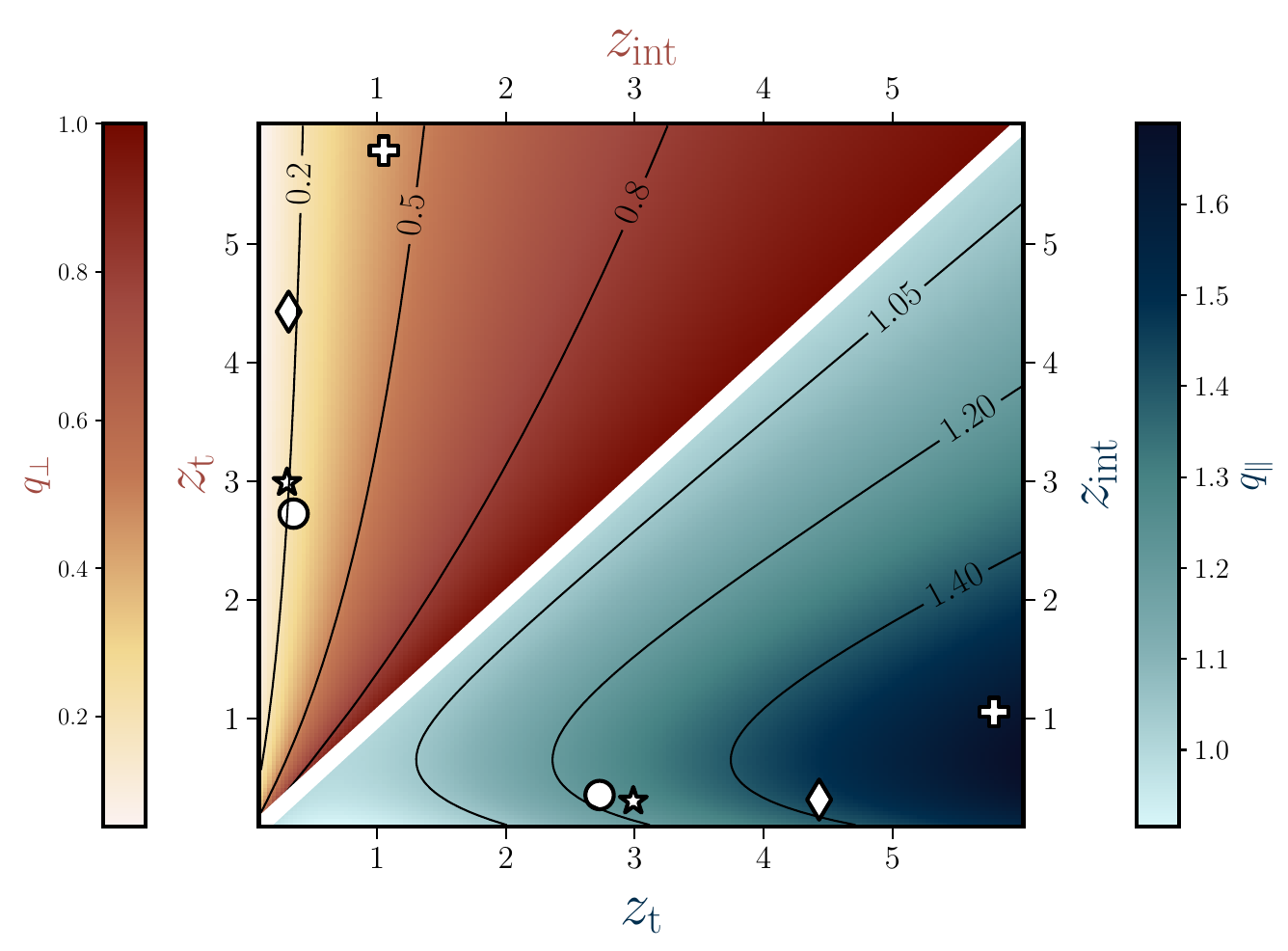}
\caption{Rescaling factors $q_\perp$ and $q_\parallel$ (red and blue, respectively, each on its own triangular panel) as functions of the redshift of the target signal and the redshift that the foreground interloper is emitted from (satisfying $z_{\rm int}<z_{\rm t}$). The markers label, for both factors, the configuration of specific experiments as indicated in the legend, where the target and interloper lines are indicated. These factors relate the \textit{true} wavenumbers in the foreground frame with the \textit{measured} ones after being projected to $z_{\rm t}$ as $k^{\rm true}_\parallel = k^{\rm meas}_\parallel/q_\parallel$ and $k^{\rm true}_\perp = k^{\rm meas}_\parallel/q_\perp$.}
\label{fig:qfactors}
\end{figure}

In Fourier space, this corresponds to $k_j^{\rm true} = k_j^{\rm meas}/q_j$. If we describe the wavevector $\kv$ in terms of its modulus $k$ and the cosine $\mu$ of the angle it subtends relative to the line-of-sight component, $\mu\equiv k_\parallel/k$, we have that the relation between the true and measured values for the interloper $i$ are~\cite{Ballinger:1996cd}
\begin{equation}
\begin{split}
    & k^{\rm true}_i = \frac{k^{\rm meas}}{q_\perp^{(i)}}\left[1+\left(\mu\right)^2\left((F_{\rm proj}^{(i)})^{-2}-1\right)	\right]^{1/2}\,\text{and} \\
    & \mu^{\rm true}_i = \frac{\mu^{\rm meas}}{F_{\rm proj}^{(i)}}\left[1+\left(\mu\right)^2\left((F_{\rm proj}^{(i)})^{-2}-1\right)	\right]^{-1/2},
\end{split}
\label{eq:scaling_kmu}
\end{equation}
where $F_{\rm proj}^{(i)} \equiv\ q_\parallel^{(i)}/q_\perp^{(i)}$. In Fig.~\ref{fig:qfactors}, we show values of $q_\perp$ and $q_\parallel$ for different sets of $z_{\rm t}$ and $z_{\rm int}$, explicitly marking some actual experiments.\footnote{In some cases, the projection effects may be very large, and this linear rescaling of separations may be inaccurate. We leave an exploration of such non-linear corrections to future work.} For simplicity, we will drop the `meas' superscript from here on out and refer to `measured' distances by default unless otherwise stated.

\subsection{Power Spectrum}
Let us denote the intensity fluctuations as $\delta I(\pmb{x})\equiv I(\pmb{x})-\langle I\rangle$, where $\langle I \rangle$ is the intensity averaged over the volume probed. These fluctuations are a biased tracer of the matter density fluctuations. In fact, for spectral lines other than 21\,cm before reionization, we can safely assume that all emission comes from galaxies.\footnote{With the exception of diffuse Lyman-$\alpha$, which may require mild tailoring of the formalism discussed in this work.}

We will focus on the power spectrum of these intensity fluctuations ---the Fourier-space analog of the two-point correlation function. This receives two kinds of contributions: the shot noise associated with the discreteness of the emitters, and the clustering.

Let us first address the clustering part. Assuming that all contributions to Eq.~\eqref{eq:Itot} come from non-overlapping volumes,\footnote{This assumption is always fulfilled when the rest-frame frequencies of the relevant contributors are far enough apart. Otherwise, the cross-correlation between them must also be taken into account in Eq.~\eqref{eq:Pclust}.} and working to linear order,\footnote{Technically, our treatment of redshift-space distortions goes beyond linear order in using a Lorentzian damping factor in Eq.~\eqref{eq:F_RSD} to mimic the effect of Fingers-of-God.} the total clustering component is given by
\begin{equation}
    P_{\rm clust}(k,\mu) = \sum_i\frac{\mathcal{B}_i^2(k_i^{\rm true},\mu_i^{\rm true},z_i)}{\mathcal{A}_v^{(i)}}P_{\rm m}(k_i^{\rm true},z_i)\,,
    \label{eq:Pclust}
\end{equation}
where the $i$ subscripts index contributions from the various lines, $\mathcal{A}_v\equiv q_\perp^2q_\parallel$ is a factor due to the isotropic volume dilation from the projection effects, $P_{\rm m}$ is the matter power spectrum in real space, and
\begin{equation}
\begin{split}
    \mathcal{B}_i \equiv & X_{\rm LI}^{(i)}(z_i)\int{\rm d}M\frac{{\rm d}n}{{\rm d}M}(z_i)\times \\ \times & L_i(M,z_i) b_h(M, z_i)F_{\rm RSD}(k_i^{\rm true},\mu_i^{\rm true},z_i)\,.
\end{split}
\end{equation}
In this last expression, ${\rm d}n/{\rm d}M$ is the halo mass function; $L$ and $b_h$ are parametrizations of the line luminosity and halo bias, respectively, that depend on both the halo mass and redshift, and $F_{\rm RSD}$ is a factor encoding redshift-space distortions:
\begin{equation}\label{eq:F_RSD}
    F_{\rm RSD}(k,\mu, z)=\frac{1+f\mu^2/b_h(z)}{1+(k\mu\sigma_{{\rm pv}}^{(i)})^2/2}\,,
\end{equation}
where $f$ is the dimensionless linear growth factor. Note that this expression includes the Kaiser effect on large scales as well as a phenomenological suppression on small scales, which we model as a Lorentzian function of the velocity dispersion $\sigma_{\rm pv}^2(z) = \int {\rm dk} P_{\rm m}(k,z)/6\pi^2$. We assume the halo mass function from Ref.~\cite{Tinker:2008ff} and the halo bias from Ref.~\cite{Tinker:2010my}.

Under the same assumption of no volume overlap between the contributors to the total line intensity, and considering only Poissonian shot noise, the shot-noise component is
\begin{equation}
    P_{\rm shot} = \sum_i \frac{\left(X_{\rm LI}^{(i)}(z_i)\right)^2}{\mathcal{A}_v^{(i)}}\int{\rm d}M\frac{{\rm d}n}{{\rm d}M}(z_i)L_i^2(M,z_i)\,.
\end{equation}

LIM experiments have limited angular and spectral resolution. Assuming single-dish observations with antennas of diameter $D$, the full width at half maximum of the beam is given by $\theta_{\rm FWHM}=1.22c/\nu_{\rm obs}D$. If we consider a Gaussian beam, this corresponds to a standard deviation $\sigma_{\rm beam}=\theta_{\rm FWHM}/\sqrt{8\log 2}$. In terms of the frequency resolution, we consider Gaussian channels with width $\delta\nu$. This corresponds to characteristic resolution limits in the directions along and transverse to the line of sight
\begin{equation}
    \sigma_\parallel = \frac{c\delta\nu(1+z)}{H(z)\nu_{\rm obs}}, \quad \text{and} \quad \sigma_\perp = D_{\rm M}(z)\sigma_{\rm beam}\,,
\end{equation}
respectively, so that the suppression associated with the beam and the spectral resolution can be expressed as~\cite{Li:2015gqa} 
\begin{equation}
     W_{\rm res}(k,\mu) = \exp\left\lbrace -k^2\left[\sigma_\perp^2(1-\mu^2)+\sigma_\parallel^2\mu^2    \right]\right\rbrace.
 \label{eq:Wk_res}
 \end{equation} 
The total power spectrum is then given by 
\begin{equation}
    P_{I}(k,\mu) = W_{\rm res}^2(k,\mu)\left(P_{\rm clust}(k,\mu) + P_{\rm shot}\right)\,.
\end{equation}
For simplicity, we consider only the monopole component of the redshift-space power spectrum, which can be obtained as $ P_0(k)=\int{\rm d}\mu P(k, \mu )/2$.

At various points in this paper we will invoke other tracers of the large-scale structure, either as proxies of the interloper fluctuations, or as fluctuation fields to be cross-correlated with our target signal. Though our formalism ---described in the next section--- is general enough to apply to any tracer of the large-scale structure, we will focus on spectroscopic galaxy surveys for simplicity. Whenever we claim that one such tracer originates from the same redshift as some emission line, we will assume that the two also cover the same volume and project them to $z_{\rm t}$ in exactly the same way. Note, however, that the redshift of the ancillary tracer is known, so that we can use separately the contributions from foreground and target volumes, contrary to the case of line interlopers.

We define galaxy number count fluctuations as $\delta_{\rm g}\equiv (n_{\rm g}(\xv)-\bar{n}_{\rm g})/\bar{n}_{\rm g}$,\footnote{More optimal fluctuation definitions can be obtained ---including for the line-intensity fluctuations--- by applying weights to minimize the variance (see e.g., Ref.~\cite{Blake:2019ddd} for a detailed example pertaining to LIM-galaxy cross-correlations).} where $n_{\rm g}$ is the local number density of galaxies and $\bar{n}_{\rm g}$ its mean. In this case, we have $\mathcal{B}_{\rm g} = b_{\rm g}F_{\rm RSD,g}$, where $b_g$ is the linear galaxy bias, and the galaxy power spectrum, including also the shot noise, is given by 
\begin{equation}
    P_{\rm g}(k,\mu) = \frac{1}{\mathcal{A}_v^{(\rm g)}}\left[\mathcal{B}_{\rm g}^2(k^{\rm true},\mu^{\rm true},z_{\rm g})P_{\rm m}(k_i^{\rm true},z_{\rm g}) + \frac{1}{\bar{n}_{\rm g}}\right]\,,
\end{equation}
where we have assumed the shot noise is Poissonian. 
On the other hand,  the cross-power spectrum with line-intensity fluctuations is
\begin{equation}
    P_{I{\rm g}} = \frac{W_{\rm res}}{\mathcal{A}_v^{(i)}}\left[\mathcal{B}_i\mathcal{B}_{\rm g}P_{\rm m} + X_{\rm LI}^{(i)}\int{\rm d}M\left.\frac{{\rm d}n}{{\rm d}M}\right\vert_{\in {\rm g}}\frac{L_i}{\bar{n}_{\rm g}}\right]\,,
\end{equation}
where we have dropped explicit dependences to ease the reading of the expression and only the source galaxies that are also featured in the galaxy survey contribute to the average luminosity density in the last integral.

\subsection{Power spectrum covariance}
Assuming white effective experiment noise, the noise power spectrum of single-dish LIM surveys is given by
\begin{equation}
    \mathcal{N}_I = \frac{\sigma_{\rm N}^2V_{\rm vox}}{N_{\rm det}}\,,
    \label{eq:Pnoise}
\end{equation}
where $\sigma_{\rm N}$ is the standard deviation of the effective noise per voxel, $V_{\rm vox}$ is the voxel volume, and $N_{\rm det}$ is the effective number of detectors. 

For analytic derivations in this work, we will only consider the Gaussian covariance of the monopole power spectra. Including the contributions from signal and instrumental noise, the covariance of the auto-power spectrum of LIM observations is given by
\begin{equation}
    \mathcal{C}_{II} = \frac{2\left(P_{I}+\mathcal{N}_I\right)^2}{N_{\rm modes}}\,,
\end{equation}
where $N_{\rm modes}$ is the number of modes per $k$-bin. 

For cross-correlations, we will assume that the noise of the LIM experiment is uncorrelated with the galaxy fluctuations, yielding
\begin{equation}
    \mathcal{C}_{I{\rm g}} = \frac{\left(P_{I}+\mathcal{N}_I\right)P_{\rm g}     +P^2_{I{\rm g}}}{N_{\rm modes}}\,,
    \label{eq:crosscovar}
\end{equation}
where $P_{\rm g}$ is the total monopole power spectrum of galaxy clustering, including shot noise.

\section{Cleaning interlopers}
\label{sec:nulling}
In this section, we describe our proposed methodology to minimize the contribution from interlopers to LIM observations using external tracers of the large-scale structure. Our approach is inspired by nulling techniques as applied to weak lensing~\cite{huterer_nulling_2005}, and  in particular to CMB lensing~\cite{Sherwin:2015baa,McCarthy:2020dgq,Qu:2022xow,baleato_lizancos_delensing_2022}.

Let us consider a simplified scenario where in addition to the target spectral line there is a single foreground line interloper, with the two lines separated enough in frequency so that there is no overlap in the volume they trace. In this limit, the intensity fluctuations are the sum of three uncorrelated fields:
\begin{equation}
    \delta I(\pmb{x}) = \delta I_{\rm int}(\pmb{x}) + \delta I_{\rm t}(\pmb{x}) + \delta_\mathcal{N}(\pmb{x})\,,
\end{equation}
where the interloper component is subject to projection effects, and $\delta_\mathcal{N}$ is the noise fluctuation, assumed to be randomly sampled from a power spectrum given by Eq.~\eqref{eq:Pnoise}. The measured power spectrum from these fluctuations is $P_{I}\equiv P_{I_{\rm t}}+P_{I_{\rm int}} + \mathcal{N}_I$. Similarly, we consider galaxy number counts from a galaxy survey covering the same volume as the interloper line emission above, leading to an overdensity field $\delta_{\rm g}^{\rm int}$.

After transforming to Fourier space, let us define a `cleaned' version of the line-intensity fluctuations as
\begin{equation}\label{eqn:field_level_cleaning}
    \delta\hat{I}(\pmb{k}) = \delta I(\pmb{k}) - \mathcal{F}(\pmb{k})\delta_{\rm g}^{\rm int}(\pmb{k})\,,
\end{equation}
where $\mathcal{F}$ is a set of weights yet to be determined. Without loss of generality, the measured power spectrum after cleaning can be modeled as
\begin{equation}
    \tilde{P}_{\hat{I}} = \tilde{P}_{I} -2 \mathcal{F} \tilde{P}_{I\rm g^{\rm int}} + \mathcal{F}^2 \tilde{P}_{{\rm g}^{\rm int}}
    \,, 
\end{equation}
where tildes denote measured quantities including noise. Then, the model that would be used for $P_I$ can be applied to the cleaned version of the map, since we know $\mathcal{F}$ exactly.

We want to improve the precision of the measurement of the power spectrum of the target line by removing contributions from line interlopers. Therefore, we choose $\mathcal{F}$ to ensure that $\delta\hat{I}$ is unbiased and the variance of its power spectrum is minimized.

The first condition is fulfilled by definition ---since $\langle\delta I\rangle=\langle \delta_{\rm g}^{\rm int}\rangle=0$ and $\mathcal{F}$ is not a statistical quantity. To fulfill the second condition we need the weights to be
\begin{equation}
    \mathcal{F}(k) = \frac{\tilde{P}_{I{\rm g}^{\rm int}}(k)}{{\tilde{P}}_{{\rm g}^{\rm int}(k)}} = \frac{\tilde{P}_{I_{\rm int}{\rm g}^{\rm int}}(k)}{{\tilde{P}}_{{\rm g}^{\rm int}(k)}}
    \,.
    \label{eq:filterweights}
\end{equation}
Note that the denominator in the filter includes noise ---which for galaxy surveys is only the shot noise already included in the definition of $P_{\rm g}$; if a different tracer including instrumental noise is used, the noise must be explicitly included in the computation of $\mathcal{F}$. Alternatively, the power spectra employed to compute the filter could be obtained from a model for the signal component only; this would null the interloper contribution exactly, but it could also introduce noise and, potentially, bias due to model mismatch (cf. Refs.~\cite{Qu:2022xow, McCarthy:2020dgq, Lizancos:2023jpo}). Instead, we prefer to obtain them from the actual data, rendering the whole procedure model independent.\footnote{This is exactly analogous to the determination of the optimal weights for CMB delensing~\cite{smith_delensing_2012, Sherwin:2015baa} and `redshift-cleaning'~\cite{Lizancos:2023jpo}. These applications have been shown to be relatively robust against inaccuracies in the determination of the filters~\cite{yu_multitracer_2017,namikawa_simons_2022}. The key is that we know exactly what filters are applied to the data, such that any filter misspecification is only likely to result in suboptimal variance reduction, but no bias~\cite{Sherwin:2015baa, yu_multitracer_2017,namikawa_simons_2022}.}

If we define the cross-correlation coefficient between two tracers as
\begin{equation}
    \rho_{ab} = \frac{{P}_{ab}}{\sqrt{({P}_{aa}+\mathcal{N}_a)({P}_{bb}+\mathcal{N}_b)}}\,,
\end{equation}
the resulting power spectrum for $\delta\hat{I}$ using the filter defined in Eq.~\eqref{eq:filterweights} is given by
\begin{equation}
\begin{split}
    \tilde{P}_{\hat{I}} & = P_{I_{\rm t}}+P_{I_{\rm int}}\left(1-\rho^2_{I_{\rm int}{\rm g}^{\rm int}}\right)+
    \mathcal{N}_I = \\
    & =\tilde{P}_{I}\left(1-\tilde{\rho}^2_{Ig^{\rm int}}\right) \,,
\end{split}
    \label{eq:autocleaned}
\end{equation}
where $\rho^2_{I_{\rm int}{\rm g^{int}}}$ in the first line is computed without the noise in the line-intensity map. The two expressions above are equivalent, since the inclusion of noise and target fluctuations in the cross-correlation coefficient reduces the value of $\tilde{\rho}_{Ig^{\rm int}}$. While the former expression provides a clearer idea of the origin of the cleaning and the reduction in the power spectrum due to a suppression of interloper clustering, the latter shows that we can predict the resulting power spectrum after the cleaning only with measurable quantities from the data. 

Note that, even if the ancillary galaxy sample traces the same underlying matter fluctuations as the interloper line, noise, instrumental resolution and nonlinear clustering will prevent the cleaning from being perfect. The correlation coefficient can differ from unity even on large scales due to shot noise from the discreteness of the galaxies, instrumental noise on the LIM observations, and contributions like non-linear galaxy bias and halo exclusion which mimic stochasticity on large scales. This has been shown explicitly in the context of LIM auto- and cross-correlations in simulations (see e.g.~\cite{MoradinezhadDizgah:2021dei, Sato-Polito:2022wiq, Obuljen:2022cjo}). 

The set of weights above can be straightforwardly extended to remove multiple interlopers, or to combine several external tracers for more effective cleaning ---more details are provided in appendix~\ref{app:multitracer} in the harmonic formalism. When the contributions from different lines trace overlapping volumes, the formalism must be extended to take into account the correlation between them. We leave the exploration of the cases that would follow this scenario for future work.

A rather unique benefit of our approach to removing interloper contamination is that we eliminate the actual realization of the contaminant present in our observations ---at least partially. Along with the contaminant, then, we also remove some of its sample variance, improving the precision of the statistics we can extract from the observations. Explicitly, the Gaussian covariance of the power spectrum of the cleaned intensity given in Eq.~\eqref{eq:autocleaned} is
\begin{equation}
    \mathcal{C}_{\hat{I}\hat{I}} = \frac{2\left[P_{I_{\rm t}}+P_{I_{\rm int}}\left(1-\rho^2_{I_{\rm int}{\rm g}^{\rm int}}\right)+\mathcal{N}_{I}\right]^2}{N_{\rm modes}}\,.
    \label{eq:filt_covar}
\end{equation}

Removing the contribution from interlopers will also be crucial to increase the detection significance of cross-correlations. This is because the auto-power spectra of both tracers contribute to the covariance of their cross-power spectrum, as shown in Eq.~\eqref{eq:crosscovar}. To see this more explicitly, consider the cross-correlation of a target line-intensity map with  galaxy number counts $n_{\rm g}^{\rm t}$ probing the same volume. The covariance of their cross-spectrum is
\begin{equation}
    \mathcal{C}_{\hat{I}{\rm g}^{\rm t}} = \frac{\left[{P}_{I_{\rm t}}+P_{I_{\rm int}}\left(1-\rho_{I_{\rm int}{\rm g}^{\rm int}}^2\right)+\mathcal{N}_I\right]{P}_{{\rm g}^{\rm t}}+P_{I_{\rm t}{\rm g}^{\rm t}}^2}{N_{\rm modes}}\,.
    \label{eq:filt_covarcross}
\end{equation}
Note that the quantity in the square brackets in Eqs.~\eqref{eq:filt_covar} and~\eqref{eq:filt_covarcross} can equally be expressed in terms of $P_{\rm I_{\rm tot}}\left(1-\rho^2_{I_{\rm tot}g^{\rm int}}\right)$, as done in Eq.~\eqref{eq:autocleaned}.

\section{Validation against simulations}
\label{sec:validation}
Before investigating the gains the cleaning offers, let us validate the analytic prediction for the power spectrum of the cleaned line-intensity map shown in Eq.~\eqref{eq:autocleaned}. We use the code  \textsc{simple}\footnote{Publicly available at \url{https://github.com/mlujnie/simple}.}~\cite{Niemeyer:2023yeu} to generate log-normal simulations of the fields of interest. \textsc{simple} is based on log-normal galaxy simulations\footnote{\url{https://bitbucket.org/komatsu5147/lognormal_galaxies/src/master/}}~\cite{Agrawal:2017khv}, to which it adds line intensities according to an input line-luminosity function. As shown in Ref.~\cite{Niemeyer:2023yeu}, \textsc{Simple} reproduces the expected power spectra and shot noise. We take the nonlinear matter power spectra as modeled by \textsc{HMCode}~\cite{Mead:2020vgs} as input for \textsc{Simple}. This nonlinear matter power spectrum partially captures the effects of nonlinear intensity fluctuations. However, since \textsc{Simple} Poisson-samples the lognormal matter fluctuations with galaxies according to a linear bias and treats each galaxy as a point source, it does not capture the emission profile and decorrelation characteristic of the one-halo term~\cite{Schaan:2021gzb}. This is a  common feature to lognormal realizations, but is not a limitation to our purposes: in terms of the formalism presented in Sec.~\ref{sec:nulling}, any effect of the one-halo term not included in the realizations can be embedded in the correlation coefficient.

We choose a setup mimicking HETDEX observations~\cite{Gebhardt:2021vfo, Niemeyer:2023yeu} to validate our analytic derivations. We consider observations in the wavelength range $\lambda\in\left[439.4,\,541.4\right]$\,nm. This range corresponds to $z_{\rm t}\in\left[2.61,\,3.45\right]$ for the target Lyman-$\alpha$ line (Ly$\alpha$, 121.6\,nm at rest frame), and $z_{\rm int}\in\left[0.18,\,0.45\right]$ for the foreground interloper [OII] line (372.7\,nm at rest frame); the \textit{center} redshifts, computed from the mean frequency, correspond to $z_{\rm t}=2.99$ and $z_{\rm int}=0.30$, respectively. In addition to the line-intensity maps, we generate log-normal realizations of galaxy number counts at the same two redshifts.

We compute the Ly$\alpha$ and [OII] luminosity functions to input to \textsc{simple} at the corresponding redshifts assuming empirical relations between the line luminosity and the star-formation rate (SFR). For the Ly$\alpha$ line we use~\cite{COMAP:2018svn}
\begin{equation}
    \frac{L_{\rm Ly\alpha}}{L_\odot} = 4.18\times 10^{8} \left(\frac{{\rm SFR}(M)}{M_{\odot} \text{yr}^{-1}}\right) f_{\rm esc}({\rm SFR}(M), z)\,,
\end{equation}
where the escape fraction is parametrized as
\begin{equation}
\begin{split}
    f_{\rm esc}&({\rm SFR}(M), z) = \left(1+e^{-1.6z+5}\right)^{-1/2} \times\\ & \times\left[0.18 + \frac{0.82}{1+0.8\left(\frac{{\rm SFR}(M)}{M_{\odot} \text{yr}^{-1}}\right)^{0.875}}\right]^2\,.
\end{split}
\end{equation}
In turn, for [OII] we use
~\cite{2017ApJ...835..273G}
\begin{equation}
    \frac{L_{\rm OII}}{L_\odot} = 1.87\times 10^7 \left(\frac{{\rm SFR}(M)}{M_{\odot} \text{yr}^{-1}}\right)10^{-0.62/2.5}\,.
\end{equation}
We assume the SFR-to-halo mass and redshift relation from Ref.~\cite{Behroozi:2012iw} along with the quenching fractions from Ref.~\cite{2019MNRAS.488.3143B}. We transform these relations to luminosity functions assuming the halo mass function from Ref.~\cite{Tinker:2008ff}, and compute the luminosity-averaged bias with the halo-bias fitting function from Ref.~\cite{Tinker:2010my}.

We generate 100 redshift-space realizations for the setup described above. We do not include instrumental noise or foreground-continuum signal and mitigation, and boost the number density for the galaxy number counts to consider a signal-dominated regime and thus a stricter test for the validation. We neglect line broadening for simplicity; we refer the reader to Refs.~\cite{COMAP:2021rny} and~\cite{Bernal:2023ovz} for a discussion about the effects of line broadening on the LIM power spectrum and 1-point statistics. The details for the procedure followed to generate each of the realizations  implementing the projection effects are described below:
\begin{figure}[th!]
\centering
\includegraphics[width=\columnwidth]{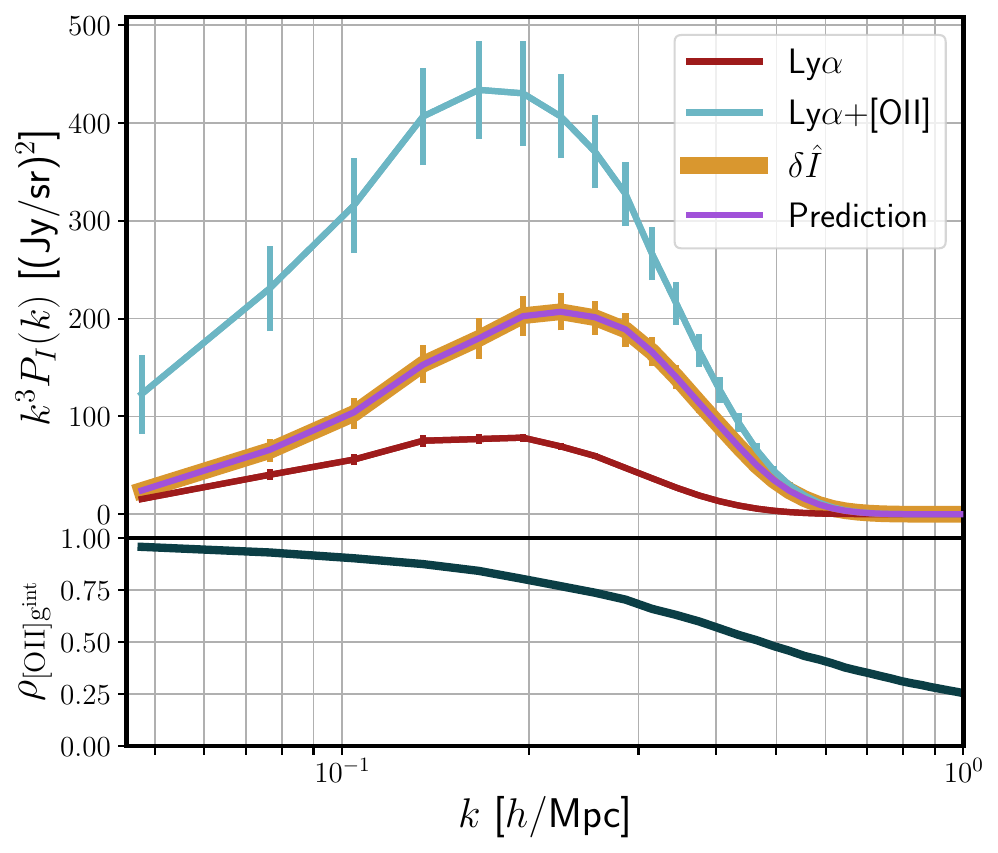}
\caption{Validation of the analytic prediction for the power spectrum of the line-intensity map cleaned using the approach proposed in this paper. Top panel: monopole of the redshift-space power spectrum of the target Ly$\alpha$ line (red) and the total, interloper-contaminated intensity [OII]$+$Ly$\alpha$ before (blue) and after cleaning (yellow). Shown errorbars are the square root of the diagonal of the corresponding covariance matrix, obtained from the 100 realizations, while the actual signal is the average over 100 realizations. The theoretical prediction for the power after cleaning (derived from the measured power spectra) lies exactly on top of the simulated result. Bottom panel: cross-correlation coefficient between the [OII] intensity fluctuations and the foreground galaxy number counts.}
\label{fig:validation}
\end{figure}
\begin{enumerate}
    \item We generate a log-normal realization of `target' Ly$\alpha$ emission at $z_{\rm t}=2.99$ over a cubic box with side length 256\,Mpc$/h$ using a grid of 128$^3$ cubic cells of side length $2\,{\rm Mpc}/h$. Using the same random seed ---to ensure the same density fluctuations underlie every realization--- we generate a log-normal realization of number counts with linear bias $b_g=2.0$ and number density 0.17\,(Mpc$/h$)$^{-3}$. 
    \item We determine the volume for the simulated foreground emission. The foreground volume at $z_{\rm int}=0.30$ that projects onto the cubic box described above is a rectangular box with side lengths $L_\perp=48.8\, {\rm Mpc}/h$ and $L_\parallel = 321.2\, {\rm Mpc}/h$. Similarly, the foreground volume that projects onto each cell of the background box is $0.38^2\times 2.51\, ({\rm Mpc}/h)^3$. However, in order to simulate all the relevant density-fluctuation modes, we consider a larger cubic box of side $321.2\, {\rm Mpc}/h$ and grid of 843$^3$ cells. 
    \item We change the random seed for the foreground distributions to avoid any correlation with the realizations described above. We generate the foreground [OII] realization and a galaxy-number-counts realization with number density 2.8 (Mpc$/h$)$^{-3}$ and bias $b_g=1.3$ using the same random seed. 
    \item We downsample the realization along the line of sight into a grid of $843^2\times 128$ cells, and save only the central $128^3$ of them. The stored (rectangular) volume of $128^3$(rectangular) cells corresponds exactly to the volume that would project onto the background box when interpreted at $z_{\rm t}$.
    \item In order to analyze only the fluctuations in the line intensities and galaxy number counts, we remove the mean of each slice along the line of sight from each map.
    \item To facilitate the analysis, we interpret all 4 fields as being contained in a nominal box of (256\,Mpc$/h)^3$. 
    \item We apply an anisotropic Gaussian filter to the line-intensity realizations, using standard deviation widths of $\sigma_\perp=4\, {\rm Mpc}/h$ and $\sigma_\parallel=6\, {\rm Mpc}/h$ to simulate limited experimental resolution.\footnote{Note that this filter does not correspond to the actual HETDEX resolution limits. More details about the HETDEX limits applied to LIM can be found in Ref.~\cite{Niemeyer:2023yeu}.} 
\end{enumerate}
Once we have all the realizations, we compute the auto- and cross-power spectra of all correlated combinations, considering also the \textit{total} line-intensity power spectrum by adding together the Ly$\alpha$ and [OII] maps. Note that since the random seed used to generate the background and foreground tracers is different, they are not correlated. 

\begin{figure*}
\centering
\includegraphics[width=\textwidth]{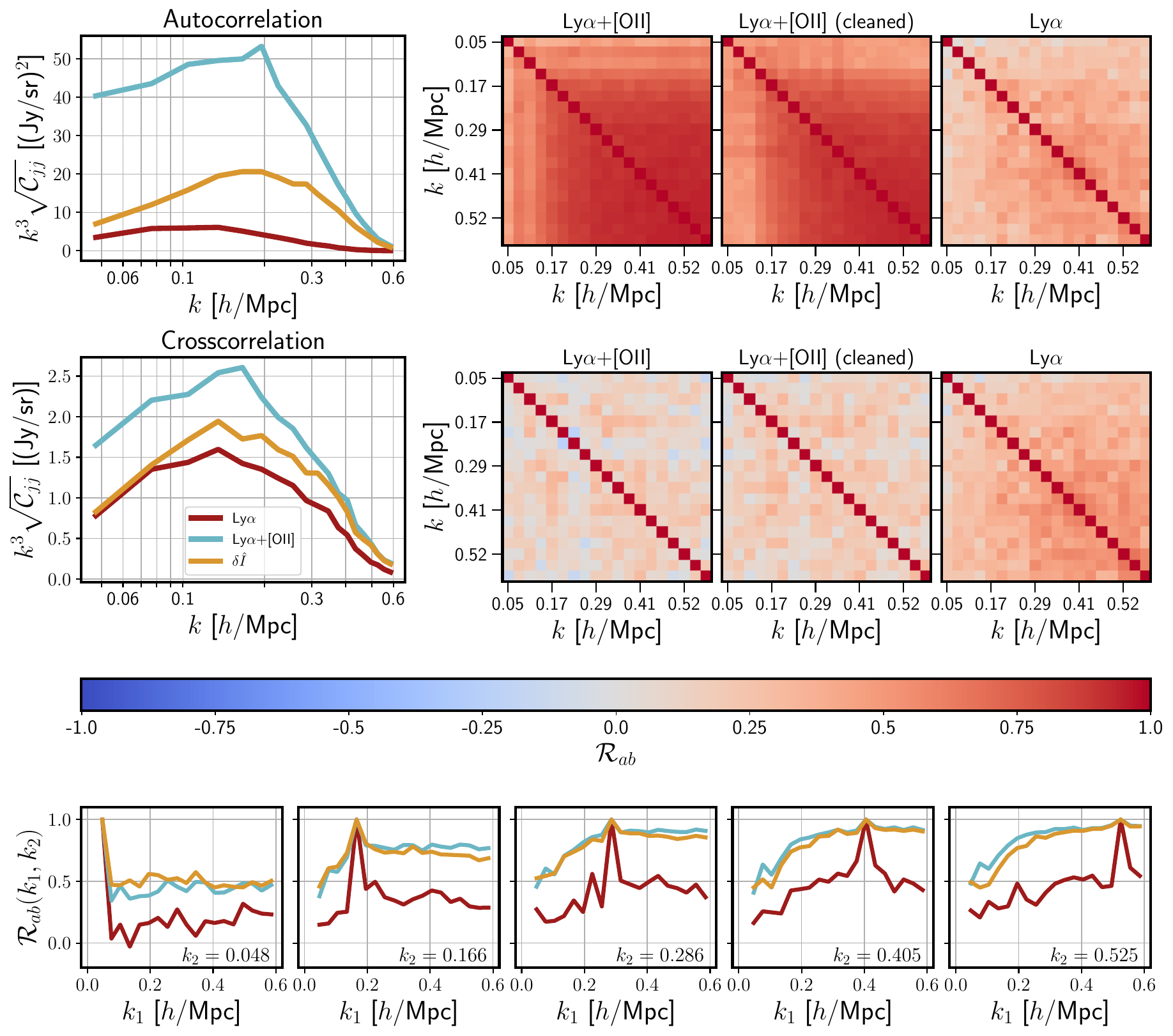}
\caption{Square-root of the variance (leftmost panels) and correlation matrices (additional panels to the right) of the LIM auto-spectrum (top) and cross-spectrum with galaxies (middle), estimated from the lognormal realizations described in Sec.~\ref{sec:validation}. For each of these, we show the case of the total intensity fluctuations (blue lines in the leftmost panels), the cleaned intensity maps (yellow lines), and the Ly$\alpha$-only intensity maps (red); the correlation matrices are labeled accordingly. We show slices of the correlation matrices in the bottom panels.}
\label{fig:covariances}
\end{figure*}

From these products we can also compute the cleaning weights $\mathcal{F}$ from Eq.~\eqref{eq:filterweights} and generate the \textit{cleaned} maps. Finally, we compute the power spectrum of the cleaned map and its cross-correlation with the background galaxies. We also compute the cross-correlation coefficients and use them to evaluate the theoretical prediction for the power spectrum of the cleaned maps, Eq.~\eqref{eq:autocleaned}, in combination with the measured power spectra of the uncleaned maps and our choice of $\mathcal{F}$. 

Figure~\ref{fig:validation} shows the measured intensity power spectra of only Ly$\alpha$ fluctuations, total intensity, and the cleaned intensity maps, as well as the theoretical prediction for the power spectrum after cleaning, derived for each realization from the measured correlations using Eq.~\eqref{eq:autocleaned}. We can see that the prediction and the measurement of the power spectrum of the cleaned map match very precisely, validating the theoretical prediction derived in Sec.~\ref{sec:nulling}. Furthermore, these results also demonstrate that our approach (partially) removes the \textit{actual} realization of the contamination, rather than an ensemble average. This is why the match between the prediction of the  power spectrum after power spectrum and the measured one from the filtered field in the simulations is so precise.

As anticipated in Sec.~\ref{sec:nulling} ---see Eqs.~\eqref{eq:filt_covar} and~\eqref{eq:filt_covarcross}---, partially removing the contributions from interlopers to the line-intensity maps does indeed significantly reduce the covariance of the auto- and cross-power spectra. We demonstrate this reduction in the error bars of Fig.~\ref{fig:validation}, and in Fig.~\ref{fig:covariances} by showing the square root of the diagonal of the covariance matrices, estimated from the 100 realizations described above, and the corresponding correlation matrices $\mathcal{R}\equiv \mathcal{C}_{ab}/\sqrt{\mathcal{C}_{aa}\mathcal{C}_{bb}}$. Notice how the cleaning significantly reduces the covariance even for the cross-power spectrum ---the amplitude of which does not receive a contribution from interlopers. 

Interestingly, Fig.~\ref{fig:covariances} reveals an additional, problematic feature arising from the interloper contribution to the measured line-intensity maps. Given the disparate values of $q_\parallel$ and $q_\perp$ that drive the projection effects of the interloper fluctuations (see Fig.~\ref{fig:qfactors}), a measured $k$ mode in the target volume corresponds to much higher $k_\perp^{\rm true}$ and smaller $k_\parallel^{\rm true}$. This results in strong mode coupling of the measured auto-power spectrum, which is manifested as strong off-diagonal covariance, as shown in the correlation matrices of Fig.~\ref{fig:covariances}. Even if the diagonal covariance is significantly reduced after cleaning, the reduction of the off-diagonal elements in the correlation matrix is much smaller. This is likely due to the fact that we used an isotropic set of filtering weights. If instead they were anisotropic ---for which they could also be tailored to minimize higher-order Legendre multipoles---, we expect that the off-diagonal correlations would be reduced further. Similarly, harmonic cleaning in narrow frequency bins effectively corresponds to anisotropic filtering weights, although the cleaning performance in this case may be limited by an increased galaxy shot noise due to the binning of the galaxy catalog. We leave the exploration of these improved filtering weights to future work.

\section{Discussion}
\label{sec:discussion}
In the previous section we validated the theoretical prediction for the power spectrum of the cleaned line-intensity map after removing the interloper using Eq.~\eqref{eqn:field_level_cleaning}. We also checked that this cleaning procedure significantly reduces the covariance of the auto- and cross-power spectra. Nonetheless, the case considered is but an example used to validate the predictions. In this section, we provide an approximate metric for the gains we can expect from removing interlopers more generally. 

Let us start by assuming the (diagonal) Gaussian covariances from Sec.~\ref{sec:nulling}, and quantifying the signal-to-noise ratio on the power spectrum of \textit{just} the target line. After removing the interlopers, this improves to
\begin{equation}
    {\rm SNR}^2_{\rm auto}=\sum_k\frac{N_{\rm modes}P_{I_{\rm t}}^2}{2\left[P_{I_{\rm t}}+P_{I_{\rm int}}\left(1-\rho^2_{I_{\rm int}{\rm g}^{\rm int}}\right)+\mathcal{N}_I\right]^2}\,,
\end{equation}
while in the case of the cross-correlation of the target signal with galaxies, we have instead
\begin{equation}
    {\rm SNR}^2_\times = \sum_k\frac{N_{\rm modes}P_{I_{\rm t}{\rm g^t}}^2}{\left[{P}_{I_{\rm t}}+P_{I_{\rm int}}\left(1-\rho_{I_{\rm int}{\rm g^t}}^2\right)+\mathcal{N}_I\right]P_{\rm g^t}+P_{I_{\rm t}{\rm g^t}}^2}\,.
\end{equation}

From the expressions above, it is clear that the signal-to-noise ratio on the target-signal components improves after cleaning. Moreover, following on from the conclusions of Eq.~\eqref{eq:autocleaned} for the power spectrum, the improvements in SNR increases as the cleaning tracer and the interloper intensity fluctuations become more correlated. If we take the ratio of the SNRs per $k$-mode before (i.e., taking $\rho_{I_{\rm int}{\rm g^{\rm int}}}=0$) and after interloper removal, we get, after minimal manipulations
\begin{equation}
    \frac{{\rm SNR}_{\rm auto}(k)}{{\rm SNR}_{\rm auto,\, no\, null.}(k)}=\frac{1+\frac{P_{I_{\rm int}}}{P_{I_{\rm t}}+\mathcal{N}}\left(1-\rho^2_{I_{\rm int}{\rm g}^{\rm int}}\right)}{{1+\frac{P_{I_{\rm int}}}{P_{I_{\rm t}}+\mathcal{N}}}}\,,
\end{equation}
and similarly for the cross-power spectrum. Note that using the Gaussian covariance as our metric to evaluate the increase in detection significance we ignore any  mode coupling in the power spectra. As we find for our example in Fig.~\ref{fig:covariances}, mode coupling can be quite large. Hence, the estimation using the expression above corresponds to idealized scenarios and shall be used just as a qualitative guide. 

\begin{figure}[t]
\centering
\includegraphics[width=\columnwidth]{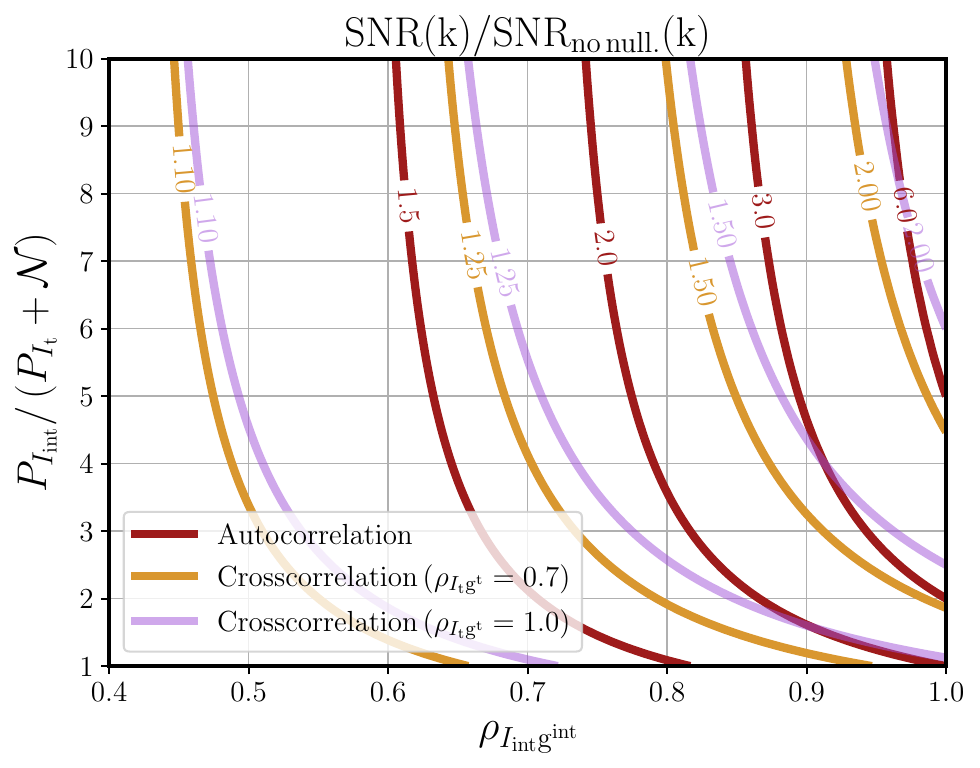}
\caption{Contour plot of the expected boost factor in the signal-to-noise ratio of auto- and cross-spectra at a given $k$ value after removing the interlopers. We show results for auto-correlation (red) and cross-correlation (assuming a cross-correlation coefficient of 0.7 and 1 in purple and orange, respectively) as functions of the ratio of the interloper and target-line power spectra and the cross-correlation coefficient between the interloper fluctuations and the tracer of the large-scale structure used to remove them.}
\label{fig:SNR}
\end{figure}

This is still useful, since we can estimate the boost factor in the detection significance for each wavenumber in the most general possible way, as a function of only the correlation coefficient between the foreground tracers and the interloper as well as of the ratio between the projected interloper power spectrum and the target-line power spectrum (including noise). We show this in Fig.~\ref{fig:SNR}. As expected, the improvement factor can be very large for auto-correlations, since interlopers contribute to the measured power spectrum and its covariance. For bright interlopers and a high correlation between their fluctuations and the tracer used to null them, we can expect improvements in the SNR of auto- and cross-correlations of up to a factors of $\sim 6$ and 2, respectively, per wavenumber. This shows how powerful this approach can be when it comes to reducing line-interloper contamination in LIM observations. 

The performance of the technique proposed in this work to remove the interlopers depends on the relative intensity between the interloper and target emissions: the brighter the interloper emission, the higher the gain after cleaning. In addition, and more importantly, the performance depends on the correlation between the interloper-intensity fluctuations and the tracer of the large-scale structure used to remove them: as this is a statistical removal, the higher the correlation, more effective is the cleaning. We can anticipate two sources of decorrelation that may limit this technique. First, shot noise from galaxy number counts reduces the correlation coefficient with other tracers. Since photometric surveys are not ideal for cross-correlations with LIM\footnote{This is because of the loss of short parallel modes in photometric surveys due to photometric redshift errors and the loss of long parallel modes in LIM observations due to continuum-contamination cleaning. Therefore, there is only a small overlap in Fourier modes between the two observed fields.} and spectroscopic galaxy surveys typically include many fewer galaxies, this may be a problem (photometric surveys with very good redshift determination may however be useful). Second, non-linear clustering effects also reduce the correlation, mostly due to stochasticity and non-linear biasing ---see e.g., Ref.~\cite{Sato-Polito:2022wiq} for an example in LIM simulations painted over analyses N-body simulation. This latter point gets aggravated by the projection effects (see Fig.~\ref{fig:qfactors}): a given $k$ measured at the background volume corresponds to much smaller scales ---and is hence more affected by nonlinear clustering--- in the perpendicular direction for the foreground fluctuations. This is why we see the interloper removal work better on large scales, where correlations are higher. Since the correlation coefficient is highly case-dependent ---depending on emission line, redshift, etc.---, we defer in-detail investigations to future work.

Throughout this work, we have neglected the one-halo term in the analytic modeling of the clustering terms of the LIM power spectrum and its cross-spectra with galaxy number-density fluctuations discussed in Sec.~\ref{sec:lim}. The one-halo term, which models small-scale non-linear clustering and the decorrelation between different tracers of the large-scale structure~\cite{Schaan:2021gzb}, dominates the clustering contribution to the power spectra at small scales. We choose not to include it in our analytic treatment in order to be consistent with the modeling implemented in the lognormal realizations and keep the discussion about the derivation of the methodology simple. This choice does not impact our conclusions or results: we only use the the analytic model to estimate the gain in the signal-to-noise ratio obtained thanks to the interloper cleaning, shown in Fig.~\ref{fig:SNR}. Any impact of the one-halo term on that expression would be captured in the context of a real analysis by a reduction in the cross-correlation coefficient, and we always express our results in terms of this quantity.

Nonetheless, the main benefit of this technique is that it is completely model independent ---it does not assume anything about the interloper, since both the ancillary map used as a tracer of its intensity fluctuations and the filter are obtained from direct measurements--- and it is straightforward to apply. The only requirements for the technique to be effective is that the ancillary map used as a proxy of the interlopers overlap with them in volume and have good redshift determination. This is fulfilled for most current and near future, as they have been planned to overlap with spectroscopic galaxy surveys  (see Fig.~9 in Ref.~\cite{Bernal:2022jap}). Crucially, since our procedure involves removing the exact realization of the interloper fluctuations, it offers a reduction in variance unmatched by similar cleaning techniques based on pure modeling approaches. Finally, estimating how much interloper signal has been removed requires only knowledge of the correlation coefficient of the intensity map with the cleaning tracer, and thus can also be obtained from the measurements.

With the exception of blind masking, other proposed techniques to remove interloper contamination rely on model assumptions. For instance, targeted masking must assume a relation between the galaxy properties and interloper intensity to select which voxels to mask; the direct modeling of interlopers or the use of spectral templates require a model of the line intensities; and the application of machine learning approaches needs accurate simulations. Considering that this method is complementary to any of the techniques listed above, its model independence can be very beneficial for an unbiased and optimal cleaning of interlopers. As an example, masking attempts to remove the brightest sources, while our interloper removal aims to eliminate statistical contributions, which may also take care of the diffuse emission. We leave the exploration of the synergies between the interloper-cleaning technique proposed here and other interloper mitigation techniques to future work. 

Note that the performance of interloper removal can be optimized by using more than one tracer of the foreground fluctuations. Appendix~\ref{app:harmonic} explains how to optimally combine multiple tracers in such a scenario. Moreover, these weights can also be used to combine samples split into various redshift bins; doing so provides added freedom to weight each shell independently and thus better match the effective redshift distribution of the combined tracer to that of the target line (discussions on this abound in the CMB lensing literature; see e.g., Refs.~\cite{Sherwin:2015baa,Qu:2022xow,namikawa_simons_2022}). 

\section{Conclusions}
\label{sec:conclusions}
In this work we have proposed to minimize the contributions from line interlopers to LIM observations by using ancillary tracers of the large-scale structure as proxies for the spatial intensity fluctuations of the contaminating signal. The only requisite for the application of our method is the availability of external, ancillary observations of tracers of the large-scale structure overlapping in volume with the interloper contribution that we aim to null. Since current and upcoming LIM surveys are being designed to overlap with galaxy surveys, such ancillary tracers should be readily available. 

The removal is achieved by subtracting the fluctuations of the ancillary tracer from the LIM observations, after having first filtered the former using a set of weights designed to minimize the variance of the power spectrum of the cleaned map. These weights can be directly calibrated from the data, so the changes in the power spectrum after cleaning can be modeled using only measurable quantities. Hence, interloper removal as presented in this work reduces the interloper contribution to the LIM power spectrum in a model-independent way without increasing the noise of the resulting map. Therefore, this is a cheap and ready-to-use technique that allows us to statistically remove line-interloper contamination. 

We have derived the optimal filter to perform the cleaning as well as a model for the power spectra of the \textit{cleaned} line-intensity fluctuations. Afterwards, we have successfully validated this prediction with log-normal simulations, and estimated the boost in detection significance for the LIM power spectrum and the cross-power spectrum with other large-scale structure tracers ---assuming a Gaussian covariance. The gain in detection significance of the power spectrum of interest depends on two factors: first, the relative brightness of the interlopers with respect to the target signal; and second, the extent of correlation between the external ancillary tracer used and the interloper fluctuations. We have found that for bright interlopers, the detection significance can improve up to a factor 6 and 2 for the LIM auto- and cross-power spectra, respectively. Note, however, that this idealized treatment ignores mode coupling, which can be quite large due to the anisotropy induced on the interloper fluctuations by projection effects.

In this work we have focused on the Legendre monopole of the redshift-space power spectrum. Given the strong anisotropies introduced in the interloper fluctuations by the projection effects, extending the cleaning to the anisotropic power spectrum may improve the performance of the nulling. For instance, the filtering weights can be designed to minimize the variance of the anisotropic power spectrum, and may also result in a reduction of the correlation between different $k$-bins. Note that improving the detection significance of the quadrupole may have a large impact on the scientific output since higher-order multipoles of the power spectrum can break degeneracies between astrophysics and cosmology~\cite{Bernal:2019jdo}. Similarly, the cleaning can be optimized for other summary statistics (e.g., higher-order correlators) by imposing the minimization of the interloper contributions at the time of deriving the optimal weights. 

We have applied the cleaning in three-dimensional Fourier space. Alternatively, the cleaning can also be applied separately to each frequency channel in harmonic space,  after which the observations can be projected to a three-dimensional Cartesian grid. In such cases, projection effects may be less of a problem and the removal of interlopers may be more tailored to each observed frequency. In addition, this approach may more efficiently reduce the mode coupling due to the projection effects of the line interloper. However, higher shot noise of the ancillary tracer may limit the performance as it reduces the correlation with the interloper signal. Moreover, we have focused on ancillary tracers with good redshift determination. Other alternatives such as photometric galaxy surveys with good redshift determination might be efficient too, but their low correlation with LIM observations after continuum foreground cleaning may severely hinder the nulling performance. Note also that the high-redshift tail of the distribution of the ancillary tracers may correlate with the target signal, which would require developing new weights ---following Ref.~\cite{Lizancos:2023jpo}. Finally,  the cleaning can be further improved employing more than one tracer of the large-scale structure ---as explored in the appendix---, and splitting the tracer into redshift bins that are weighted separately in order to better match the redshift distribution of the target.

In summary, this work proposes a new way to remove line-interloper emission from LIM observations that is also highly complementary to previous mitigation strategies. While other cleaning techniques focus on the brightest sources and are very model dependent, our proposed removal approach eliminates contributions statistically and in a model independent way. Moreover, it can potentially deal with diffuse emission on large scales too. Therefore, we expect this approach to perform especially well in combination with other techniques. In a similar fashion, this approach can easily be extended to deal with continuum foregrounds both for LIM and wide-band observations.

In light of the expected sensitivity of forthcoming LIM surveys and their extensive overlap with spectroscopic galaxy surveys, we think it is worthwhile to further develop this technique as required by specific scenarios to deliver on the vast scientific promise of line-intensity mapping. 

\begin{acknowledgments}
The authors would like to thank the Centro de Ciencias de Benasque Pedro Pascual, Spain, and the organizers of the Understanding Cosmological Observations workshop, where work on this paper first started. We also thank Yun-Ting Cheng and Mathilde Van Cuyck for comments on the manuscript of this work. JLB acknowledges funding from the Ramón y Cajal Grant RYC2021-033191-I, financed by MCIN/AEI/10.13039/501100011033 and by
the European Union “NextGenerationEU”/PRTR, as well as the project UC-LIME (PID2022-140670NA-I00), financed by MCIN/AEI/ 10.13039/501100011033/FEDER, UE. The authors thank the computer resources provided by the Spanish Supercomputing Network (RES) node at Universidad de Cantabria and the Institute of Physics of Cantabria (IFCA-CSIC).
\end{acknowledgments}

\appendix
\section{Harmonic-space formalism}
\label{app:harmonic}
We have focused on three-dimensional line-intensity maps, as is customary in LIM analyses. Alternatively, it is also possible to null interlopers \textit{before} projecting the observations to three-dimensional space, in the context of a radial-angular split that is better suited to the basis of the observations and later apply the projection. 

Suppose an observed field is given by $X(\n, \nu_{\rm obs})$. The map of diffuse emission in a given channel ---or group of channels--- can be binned into a pixelized map and transformed to spherical harmonic space to obtain the spherical harmonic coefficients $X_{\ell m}(\nu_{\rm obs})$. In the case of discrete sources such as galaxies, the pixelization step can be bypassed and replaced with a direct spherical harmonic transform in order to avoid issues such as aliasing~\cite{baleato_lizancos_harmonic_2024}. In this radial-angular formalism, the connection to the CMB lensing nulling/delensing literature is readily apparent (see e.g., Refs.~\cite{smith_delensing_2012, Sherwin:2015baa, baleato_lizancos_delensing_2022, McCarthy:2020dgq, Qu:2022xow, baleato_lizancos_impact_2021, Lizancos:2023jpo}). Hereinafter, we consider each $\nu_{\rm obs}$ separately and therefore drop the explicit notation.

The cleaning operation then proceeds by replacing equation~\eqref{eqn:field_level_cleaning} with its harmonic-space equivalent
\begin{equation}
    \delta\hat{I}_{\ell m} = \delta I_{\ell m} - \mathcal{F}_{\ell}\delta_{{\rm g},\,\ell m}^{\rm int}\,.
\end{equation}
In this basis, the filter takes the form
\begin{equation}
    \mathcal{F}_{\ell} = \frac{C_{\ell}^{I{\rm g^{int}}}}{C_{\ell}^{\rm g^{int}}} = \frac{C_{\ell}^{I_{\rm int}{\rm g^{int}}}}{C_{\ell}^{\rm g^{int}}}\,,
    \label{eqn:vanilla_filter}
\end{equation}
which now involves angular power spectra $C_{\ell}$ of the fluctuations in the various different channels including all sources of noise. Once this cleaning has been performed, the inverse harmonic transform can be applied to $\delta\hat{I}_{\ell m}(\nu_{\rm obs})$ to build the three-dimensional map. 

While applying interloper removal in harmonic space may be more convenient to avoid artifacts due to the projection onto a Cartesian grid~\cite{Cunnington:2023aou} and reduce the mode coupling introduced by the strong projection effects of the interlopers, the larger shot noise incurred when using narrow frequency channels may limit the performance of the cleaning. In contrast, using wider frequency bands to reduce the shot noise may hinder the cleaning due to the loss of radial modes. The next appendix will use the harmonic-space formalism to adapt more easily the results in the CMB lensing context, but they can be straightforwardly adapted to three-dimensional Fourier space.

\section{Interloper removal with multiple tracers}
\label{app:multitracer}
The combination of multiple tracers of the large-scale structure to improve CMB lensing nulling or delensing has been already proposed. In particular, Ref.~\cite{Sherwin:2015baa} includes a description of the filters that maximize the cross-correlation coefficient between the co-added tracer and the CMB lensing potential, showing that these are also the ones that minimize the variance of CMB $B$-mode polarization after delensing. Refs.~\cite{yu_multitracer_2017,namikawa_simons_2022} then showed that the procedure is relatively robust against inaccuracies in the determination of the filters.

The optimal cleaned map coadding information from several tracers takes the form
\begin{equation}
    \delta\hat{I}_{\ell m} = \delta I_{\ell m} - \sum_p c^{(p)}_{\ell} B^{(p)}_{\ell m}\,
\end{equation}
where $p$ indexes the various tracers $B$. The weights that maximize the cross-correlation coefficient between the cleaning tracer and the interloper line emission ---hence minimizing the contribution of interlopers in the cleaned map, see Eq.~\eqref{eq:autocleaned}--- are~\cite{Sherwin:2015baa} 
\begin{equation}
    c^{(p)}_{\ell} = \sum_{j}\rho^{jI_{\rm int}}_\ell\left(\rho^{-1}\right)^{pj}_{\ell}\sqrt{\frac{C_{\ell}^{I_{\rm{int}}}}{C_{\ell}^{B^{(p)}} + \mathcal{N}_{\ell}^{B^{(p)}}}}
    \,, 
    \label{eqn:multitracer_weights}
\end{equation}
where $\left(\rho^{-1}\right)^{pj}_{\ell}$ is the $pj$ element of the inverse of the matrix composed of the cross-correlation coefficients among all ancillary tracers for each multipole. Qualitatively, on a given angular scale and frequency channel, this scheme weights more heavily the tracers that best correlate with the interloper line. A similar expression can be derived for three-dimensional statistics in Fourier space, so that they can be applied to improve the formalism described in the main text, if more than one tracer is available.

It is worth noting that the dependence on $C_{\ell}^{I_{\rm{int}}}$ in Eq.~\eqref{eqn:multitracer_weights} is rather benign. Though not measurable directly, this component can typically be modeled. Moreover, the performance is moderately robust againts model misspecification~\cite{Sherwin:2015baa, yu_multitracer_2017,namikawa_simons_2022}, while most likely not introducing any bias in the cleaned map. This is because the filters that were applied to the data are known exactly, however suboptimal they may be.

\bibliography{Refs}
\bibliographystyle{utcaps}

\end{document}